\documentclass[twocolumn,multicolumn,aps,prb,showpacs]{revtex4}

\usepackage{graphicx}
\usepackage{dcolumn}
\usepackage{bm}
\usepackage{color}
\usepackage{latexsym}
\usepackage{epsfig}

\begin{document}

\preprint{APS/123-QED}

\title{ Thermodynamics of a heavy ion-irradiated superconductor: the zero-field transition}
%
%
%
\author{C.J. van der Beek}
\affiliation{
Laboratoire des Solides Irradi\'{e}s, CNRS-UMR 7642 \& 
CEA/DSM/DRECAM, Ecole Polytechnique, 91128 Palaiseau cedex, France}
%
%

\author{Thierry Klein, Ren\'{e} Brusetti}
\affiliation{Institut N\'{e}el,  B.P. 166, 38042 Grenoble cedex 9, France
}%
\author{Christophe Marcenat}
\affiliation{D\'{e}partement de Recherche Fondamentale sur la Mati\`{e}re Condens\'{e}e /SPSMS/LaTEQS, Commissariat \`{a} l'Energie Atomique, 17 Avenue des Martyrs, 38054 Grenoble cedex 9, France
}

\author{Mats Wallin}
\affiliation{Department of Theoretical Physics, KTH, AlbaNova University Center, SE-106 91 Stockholm, Sweden}

\author{S. Teitel}
\affiliation{Department of Physics and Astronomy, University of 
Rochester, Rochester, New York 14627, U.S.A.}

\author{Hans Weber}
\affiliation{Division of Physics, Lule\aa \hspace{1pt} University of Technology, 
SE-971 87 Lule\aa, Sweden}

\date{\today}

\begin{abstract}
  Specific heat measurements show that the introduction of amorphous columnar
  defects considerably affects the transition from the normal to 
  the superconducting state in {\em zero} magnetic field.   
  Experimental 
  results are compared to numerical simulations of the 3D XY model for 
  both the pure system and the system containing random columnar 
  disorder.   The numerics reproduce the salient features of experiment, 
  showing in particular that the specific heat peak changes from cusp-like to smoothly
  rounded when columnar defects are added.  By considering the specific heat
  critical exponent $\alpha$, we argue that such behavior is consistent with
  recent numerical work\cite{Vestergren2004i} showing that the introduction of
  columnar defects changes the universality class of the transition.
\end{abstract}

\maketitle

An enormous amount of attention has been paid in recent years to the 
effect of amorphous columnar defects on the superconducting transition 
in a magnetic field.\cite{nelson93,Feigelman93,Blatter94,KrusinElbaum94,Larkin95,Samoilov96,Nakielski99}
Much less work has been done on the transition in zero magnetic 
field, the only report to our knowledge being the measurement of the 
microwave conductance transverse to the columnar 
defects.\cite{Nakielski99} Nevertheless, the transition in zero 
magnetic field merits attention in its own right.\cite{Vestergren2004i}  It is 
expected that disorder will be a relevant perturbation, and change the 
universality class of a phase transition, whenever $2-d^{*}\nu 
>0$ (modified Harris criterion\cite{Wallin1994}). Here $d^{*}$ is the number of dimensions in 
which the system is disordered, and $\nu$ is the usual correlation length 
critical exponent. In the absence of columnar defects, we expect the 
superconducting phase transition to fall in the universality class of the 
three dimensional (3D) XY model, so that $\nu=0.6717(1)$.\cite{Campostrini} In 
the case of random point disorder, $d^{*} = 3$, so that $2-d^{*}\nu \approx -0.015 < 0$,
and the disorder is irrelevant.
In the case of columnar disorder, however, $d^{*}
= 2$, so that $2-d^{*}\nu\approx 0.66 > 0$, and disorder should be relevant and
drive the system to a new universality class. Note that the stability of the new
disordered critical point with respect to the modified Harris criterion requires
that the new correlation length critical exponent satisfies  $\nu > 
1$.\cite{Vestergren2004i} Recent simulations of a columnar disordered XY model in Ref.~\onlinecite{Vestergren2004i} supported such expectations, finding a phase transition 
with anisotropic scaling and a value for the critical exponent $\nu\approx 1.2$.

It is the purpose of this report to test these ideas on a real 
superconducting system. For this, we have chosen to measure the 
specific heat of optimally doped, single crystalline 
YBa$_{2}$Cu$_{3}$O$_{7-\delta}$, both without and with columnar defects. 
It has been shown previously\cite{Overend94,Marcenat95} that the specific heat at the superconducting 
transition of pristine YBa$_{2}$Cu$_{3}$O$_{7-\delta}$ is consistent
with that of the 3D XY model, with a small negative specific heat exponent $\alpha \approx 
-0.015$.\cite{Campostrini} It turns out that this exponent is
considerably modified by the introduction of the columnar defects.

Experiments were done on two YBa$_{2}$Cu$_{3}$O$_{7-\delta}$ single 
crystals, cut from the same piece. The original crystal was grown by the flux 
method in Au crucibles, and subsequently annealed in oxygen 
in Pt tubes.\cite{Holtzberg} One crystal, which contained a single 
family of twin boundaries separated by a distance of approximately 10 $\mu$m,  
was irradiated with 5.8 GeV Pb ions at the Grand Acc\'{e}l\'{e}rateur 
National d'Ions Lourds (GANIL) in Caen, France, to a fluence of $1 
\times 10^{11}$ ions cm$^{-2}$. The ion beam was directed parallel to 
the $c$-axis; each ion impact created an amorphous columnar track of 
radius $\approx 3.5$ nm. The second, untwinned crystal, was not 
irradiated, but kept as the pristine 
reference sample. Specific heat measurements have been performed in the 
absence of an applied magnetic field, using the same measurement 
technique employed in Ref.~\onlinecite{Marcenat2003}.

\begin{figure}[tbp]
    \centerline{\epsfxsize 8.5cm \epsfbox{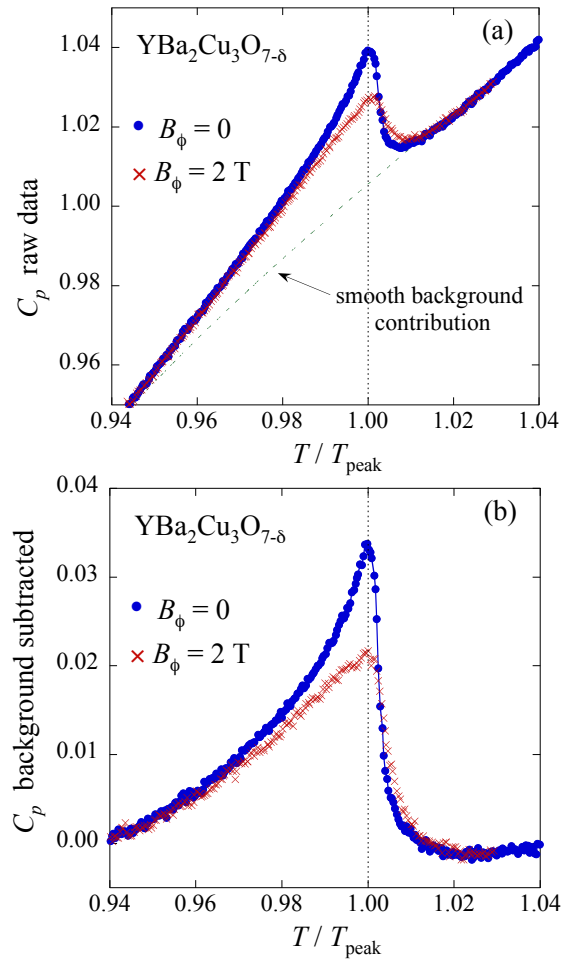}}
    \caption{  
(a) Raw specific heat data (in arbitrary units) taken on a pristine YBa$_{2}$Cu$_{3}$O$_{7-\delta}$ crystal ($\bullet$), and on a crystal with a columnar defect density $n_{d} = 1 \times 10^{11}$ cm$^{-2}$, \em i.e. \rm a matching field $B_{\phi} = \Phi_{0}n_{d} = 2$ T  ($\times$). In order to compare the two data sets, the temperature has been rescaled by the value $T_{\rm peak}$ at which the specific heat peak occurs. The dashed line represents the smooth background contribution to the specific heat, which is estimated using a third order polynomial fit. This background is subtracted in (b) in order to emphasize the shape of the specific heat peak due to the superconducting transition.  }
    \label{f1}
\end{figure}

In Fig.~\ref{f1}a we show our raw data for the specific heat of both the 
pristine and irradiated crystals.  The temperature axis has been scaled by the
value of the temperature $T_{\rm peak}$ at the specific heat peak, in order
to better compare the two samples ($T_{\rm peak}=93.1$~K for the pristine sample,
while $T_{\rm peak}= 92.1$~K for the irradiated sample).  We see clear specific heat anomalies, 
signatures of the superconducting phase transition, superimposed on
a smooth increasing background.  In the pristine sample, the amplitude of 
the specific heat anomaly was of the order of 4\% of the total specific heat, 
attesting to its very high quality. Fitting the smooth background to a cubic polynomial
(dashed line in Fig.~\ref{f1}a) we subtract this background from the data in
Fig.~\ref{f1}b, in order to better emphasize the shape of the anomaly.  For the
pristine sample, we see the typical ``$\lambda$"--cusp shape expected for 
the $H = 0$ superconducting transition in the presence of strong thermal fluctuations. 
The introduction of the amorphous columnar defects {\em 
reduces} the absolute temperature at which the specific heat 
maximum occurs. Note that a lowering of the critical temperature after heavy ion
irradiation may occur as a result of ``self-doping'' of the intercolumn
space by O ions expelled from the tracks;\cite{Pomar2000,MingLi2002} 
however, no such effect was reported for YBa$_{2}$Cu$_{3}$O$_{7-\delta}$. Another
possibility is that the columns reduce the average $T_{c}$ at which long range 
superconducting order can set in.\cite{vdBeek96,Braverman2002}     
The resulting specific heat-curve after heavy-ion bombardment shows notable 
differences with respect to the curve before irradiation. Most specifically, 
we find that the shape of the maximum is now
smoothly rounded rather than the sharp cusp seen in the pristine sample. Following the 
suggestion\cite{Vestergren2004i} that the introduction of 
columnar defects changes the universality class of the superconducting 
transition in zero field, we propose that the associated change of critical
exponents is at the origin of the markedly different shape of the
specific heat peak before and after irradiation.

\begin{figure}[tbp]
    \centerline{\epsfxsize 8.5cm \epsfbox{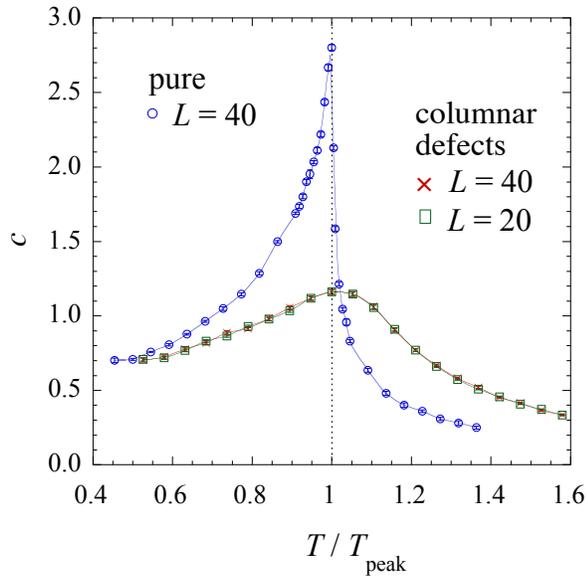}}
    \caption{Monte-Carlo calculation of the specific heat using the 3D 
    XY model, for a pure (pristine) superconductor ($\circ$), and a 
    superconductor containing columnar disorder. For the pure case,
    a system of size $40^3$ was used, while for the disordered case the
    calculation was done for two system sizes, with $L=40$ ($\times$) and 
    $L=20$ ($\Box$) grid points in the direction perpendicular to the 
    anisotropy axis, respectively; parallel to the anisotropy axis, $L_z=0.5L^{1.3}$.
    The temperature axis has been rescaled by the value $T_{\rm peak}$ at which the specific heat peak occurs.}
    \label{f2}
\end{figure}

The presence of columnar defects implies that, even in zero 
magnetic field, critical scaling of physical quantities may be 
anisotropic: the correlation length parallel to the columns, $\xi_{z}$, diverges 
as a different power of the reduced temperature $t\equiv (T-T_{c})/T_{c}$ 
than does the correlation length in the transverse direction, 
$\xi_{\perp}$. This defines the anisotropy exponent $\zeta$,
\begin{equation}
    \xi_{z} \sim \xi_{\perp}^{\zeta} .
    \end{equation}
Defining the correlation length exponent $\nu$ in the usual way, 
$\xi_{\perp} \sim |t|^{-\nu}$, the singular part of the free 
energy density will scale as 
\begin{equation}
    f(T) \sim (\xi_{z}\xi_{\perp}^{2})^{-1} \sim 
    \xi_{\perp}^{-2-\zeta} \sim |t|^{\nu(2+\zeta)}.
    \end{equation}
As a consequence, the specific heat per unit volume,  $c$, will scale as 
\begin{equation}
    c \sim \frac{\partial^{2} f}{\partial t^{2}} \sim 
    |t|^{\nu(2+\zeta)-2} \equiv |t|^{-\alpha}.
    \end{equation}
Thus, the specific heat exponent in the anisotropic case is $\alpha = 
2 - \nu(2+\zeta)$. For the pure system, {\em i.e.} a superconductor without 
columnar defects, the anisotropy exponent $\zeta = 1$, and $\nu \gtrsim 
\frac{2}{3}$. The specific heat exponent $\alpha$ is small and slightly negative.
In contrast, the calculation for the XY model of a superconductor
with columnar defects of Ref.~\onlinecite{Vestergren2004i} 
found values $\zeta \approx 1.3$ and 
$\nu \approx 1.2$. In this case, $\alpha \approx -2.0$, and is thus much more 
strongly negative. 

Because $\alpha$ is negative for both cases, the specific heat does 
not diverge at the transition. However, an interesting difference is 
seen if we consider the temperature derivative of the specific heat,
\begin{equation}
    \frac{dc}{dt} \sim |t|^{-\alpha - 1}\enspace.
    \end{equation}
For the pure superconductor, $-\alpha - 1\approx -0.985$ is negative. Therefore, 
the {\em slope} of the specific heat diverges at $T_{c}$, giving rise 
to the familiar cusp observed in Fig.~\ref{f1} for the pristine sample. For the 
superconductor with columnar defects however, $-\alpha - 1 \approx 
1.0$ is {\em positive}, so the slope of $c$ does {\em not} diverge. There is 
no sharp cusp, as is indeed observed experimentally.\cite{note}

The above discussion considers only the singular part of the specific 
heat that comes from the large length scale critical fluctuations. It 
disregards the smooth {\em non-singular} contribution that comes from 
the non-critical short length scale fluctuations. Most likely, this non-singular 
part of the specific heat has some non-zero temperature derivative 
at $T_{c}$. For the unirradiated superconductor, this temperature 
derivative is much smaller than the diverging slope of the singular 
part, and can therefore be disregarded. Thus, the non-singular 
contribution is unlikely to affect the shape of the $c$ curve very 
much. In the presence of columnar defects, however, the slope of the 
singular part vanishes at $T_{c}$. Then, the temperature derivative of 
the specific heat at $T_{c}$ is determined by the non-singular part. 
Since the slope of $c$ is non-zero and smooth at $T_{c}$, the 
{\em maximum} of $c$ is no longer located at the critical temperature,
and the shape of the peak is now a smoothly curved maximum. 
As we discuss below, this feature too is found in the experimental data.

To illustrate the comparison further, we have 
carried out Monte Carlo simulations to numerically compute $c$ 
for the pure system, and also for the system with 
columnar disorder, using the same XY model and 
random distribution for the columnar disorder as defined in Sec. IIA of Ref.~\onlinecite{Vestergren2004i}.
For the pure case, we use a system size of $40^3$ grid sites. 
For the case with columnar disorder we used two different system sizes.  Taking
$L=L_x=L_y$ as the system length transverse to the direction of the columnar
defects, and using $L_z=0.5L^\zeta$ ($\zeta=1.3$) consistent with anisotropic scaling,
we considered sizes
$L = 20$, averaged over 125 different realizations of the random 
disorder, and $L=40$, averaged over 66 different realizations of disorder.  
Our results are shown in Fig.~\ref{f2}. It is clearly seen that the pure case 
displays a peak with the familiar ``$\lambda$''--cusp, whereas the case with columnar 
disorder displays a smoothly rounded peak. We observe essentially no size 
dependence in our results for the case with columnar disorder. 
Our numerical results thus reproduce the main qualitative features  found experimentally 
for single-crystalline YBa$_{2}$Cu$_{3}$O$_{7-\delta}$. 
Difference in detail between the experimental and model curves for the case with 
columnar defects most likely results from the simplicity of the numerical model
as well as differences in the effective strength of the disorder. 
For the simulations, a particularly strong disorder was chosen so as 
to reach the asymptotic limit even with the small system sizes under 
scrutiny. Presumably, this explains why the model curve is even 
rounder than the experimental one. 

Another feature of the numerical results is that, in agreement with the argument
above, the system with columnar disorder is found to have a specific heat peak that lies
{\em below} the critical temperature, with $( T_{c} - T_{\rm peak} )/ T_{c} =
0.07$. In order to compare this result with experiment, the critical temperature
of the YBa$_{2}$Cu$_{3}$O$_{7-\delta}$ crystals is deduced from the functional
dependence of the thermodynamic properties of the superconductor on a
relevant ``scaling'' parameter. In Ref.~\onlinecite{vdBeek2005}, it was shown that, for
magnetic fields $B$ larger than 1~T, the field and temperature dependence
of the magnetization of the crystals under study can be described by a unique
(Ginzburg-Landau Lowest Landau Level, GL-LLL) relation,
$M / (\vartheta b )^{2/3} \propto F( Q )$, with $Q \equiv
(1-b)(1-\vartheta^{2})^{1/3}(\vartheta b)^{-2/3}$, where $\vartheta \equiv T /
T_{c}$ and $b \equiv B/B_{c2}(T)$; here $T_c$ is the zero field critical
temperature and  $B_{c2}(T)$ the upper critical field.
For the unirradiated crystal, this relation gives a critical
temperature $T_c= 93.1$~K, which coincides to within 0.1 K with the position of the 
zero field specific heat
maximum. For the irradiated crystal, a similar analysis gives, within the accuracy of
the fit, the same critical
temperature $T_c=93.1$~K, which now lies above the observed specific
heat maximum at 92.1~K.  The experimental data thus show a separation 
$( T_{c} - T_{\rm peak} )/T_{c} = 0.01$, in the same direction as the numerical
results, but smaller in magnitude, again presumably due to the larger disorder
strength used in the numerical model calculations.

In summary, we have measured the specific heat of pristine and 
heavy-ion irradiated single crystalline YBa$_{2}$Cu$_{3}$O$_{7-\delta}$ 
in zero applied magnetic field. The results were compared to
Monte Carlo simulations of the 3D XY model for both the pure case and 
the case with columnar disorder. Both experiment and numerics show a 
drastic influence of the columnar defects on the shape of the specific heat 
anomaly at the superconducting-to-normal transition.
The overall features of the 
specific heat anomaly are well explained by the critical exponents 
obtained from the numerical calculations, suggesting that the 
introduction of the columnar defects does indeed change the  universality class
of the transition.

We are very grateful to F. Holtzberg for providing the YBa$_{2}$Cu$_{3}$O$_{7-\delta}$ crystals.
The work of S.~T. is supported by US Department of Energy grant DE-FG02-06ER46298.
The work of M.~W. is supported by the Swedish Research Council (VR) and
the Swedish National Infrastructure for Computing (SNIC). H.~W. acknowledges support
from Swedish Research Council Contract No. 621-2001-2545.


\begin{thebibliography}{99}
    
\bibitem{nelson93}D.R. Nelson and V.M. Vinokur, Phys. Rev. Lett. 
{\bf 68}, 2398 (1992); {\em ibid.}, Phys. Rev. B {\bf 48}, 13060 (1993).

\bibitem{Blatter94} G. Blatter, M.V. Feigel'man, V.B. Geshkenbein,
    A.I. Larkin and V.M. Vinokur, Rev. Mod. Phys. 66, 1125, Ch. IX 
    (1994), and references therein.
    
\bibitem{Feigelman93} M.V. Feigel'man, V.B. Geshkenbein, L.B. Ioffe, 
and A.I. Larkin, Phys. Rev. B {\bf 48}, 16641 (1993).
   
\bibitem{KrusinElbaum94} L. Krusin-Elbaum, L. Civale, G. Blatter, A. D. Marwick, 
    F. Holtzberg, and C. Feild, Phys. Rev. Lett. {\bf 72}, 1914 (1994); 
    A.V. Samoilov and M. Konczykowski, Phys. Rev. Lett. {\bf 75}, 186 (1995). 
    L. Krusin-Elbaum, G. Blatter, and L. Civale, Phys. Rev. Lett. {\bf 75}, 187 (1995).

\bibitem{Larkin95} A.I. Larkin and V.M. Vinokur, Phys. Rev. Lett. {\bf 75}, 4666 (1995). 
    
\bibitem{Samoilov96} A.V. Samoilov, M.V. Feigel'man, M.
    Konczykowski and F. Holtzberg, Phys. Rev. Lett. {\bf 76}, 2798 (1996); 
    M. Konczykowski and A.V. Samoilov, Phys. Rev. Lett. {\bf 78}, 1830 
    (1997).

\bibitem{Nakielski99} G. Nakielski, A. Rickertsen, T. Steinborn, J. 
Wiesner, G. Wirth, A.G.M. Jansen, and J. K\"{o}tzler, in {\em Advances 
in solid State Physics 39}, edited by B. Kramer (Vieweg \& Sohn, 
Braunschweig/Wiesbaden, 1999), p.371 (also : con-mat/9904279).

\bibitem{Vestergren2004i} A. Vestergren, M. Wallin, S. Teitel, H. 
Weber, Phys. Rev. B {\bf 70}, 054508 (2004).

\bibitem{Wallin1994} M. Wallin, E.~S. S{\o}rensen, S.~M. Girvin, and A.~P. Young, Phys. Rev. B {\bf 49}, 12115 (1994).

\bibitem{Campostrini} M. Campostrini, M. Hasenbusch, A. Pelissetto,
and E. Vicari, Phys. Rev. B 74, 144506 (2006).

\bibitem{Overend94} N. Overend, M.A. Howson, and I.D. Lawrie, Phys. 
Rev. Lett. {\bf 72}, 3238 (1994).

\bibitem{Marcenat95} C. Marcenat, R. Calemczuk, and A. Carrington, 
in {\em Coherence in High Temperature Superconductors}, ed. G. 
Deutscher and A. Revcolevschi  (World Scientific, Singapore 1996) p. 101.

\bibitem{Holtzberg} F. Holtzberg and C. Feild, Eur. J. Solid State Inorg. Chem.
{\bf 27}, 107 (1990).

\bibitem{Marcenat2003} C. Marcenat, S. Blanchard, J. Marcus, L. M.
Paulius, C. J. van der Beek, M. Konczykowski, and T. Klein, Phys. Rev. Lett. {\bf 90}, 037004 (2003).

\bibitem{Pomar2000} A. Pomar, Z. Konstantinovic, L. Martel, Z. Z. Li, and H. Raffy,
Phys. Rev. Lett. {\bf 85}, 2809 (2000).

\bibitem{MingLi2002} M. Li, C.J. van der Beek, M. Konczykowski, H.W. Zandbergen, and P.H. Kes,
Phys. Rev. B {\bf 66}, 014535 (2002).

\bibitem{vdBeek96} C.J. van der Beek, M. Konczykowski, T.W. Li, P.H. Kes, and W. Benoit,
Phys. Rev. B {\bf 54}, R792-R795 (1996) 

\bibitem{Braverman2002} G.M. Braverman, S.A. Gredeskul, and Y. Avishai,
Phys. Rev. B {\bf 65}, 054512 (2002).


\bibitem{note} Note that the specific heat derivative in general will have the form $dc/dt = A_\pm |t|^{-\alpha-1}$, where the coefficient $A_+$ for $t>0$ is different from
the coefficient $A_-$ for $t<0$.  Thus, although for $\alpha\approx -2.0$ the first temperature derivative of $c$ is continuous at $T_c$, the second temperature derivative  should jump discontinuously at $T_c$.

\bibitem{vdBeek2005} C.J. van der Beek, M. Konczykowski, L. Fruchter, R.
Brusetti, Thierry Klein, Jacques Marcus, C. Marcenat, Phys. Rev. B {\bf
72}, 214504 (2005).

\end{thebibliography}
\end{document}